\documentclass[a4paper,12pt]{extarticle}
\usepackage{parskip}
\usepackage{titling}
\usepackage{authblk}
\usepackage[a4paper, left=20mm, top=20mm, right=20mm, bottom=20mm]{geometry}
\usepackage[T2A]{fontenc}			
\usepackage[utf8]{inputenc}			
\usepackage{graphicx}
\usepackage{float}
\usepackage{amsthm}
\usepackage{amsmath}
\usepackage{amssymb}
\usepackage{color}
\usepackage[font={small}]{caption}
\begin{document}
\title{Mechanical orientation of fine magnetic particles in powders by an external magnetic field: simulation-based optimization}
\author{Sergey~Erokhin, Dmitry~Berkov}
\affil{General Numerics Research Lab, Moritz-von-Rohr-Stra{\ss}e~1A, D-07745 Jena, Germany}

\maketitle
\begin{abstract}
We present a numerical algorithm for predicting the optimal conditions for the effective alignment of magnetic particles in dense powders during the compactization process using an externally applied field. This task is especially important for the permanent magnets development due to the fact that alignment of anisotropy axes of nanocomposite grains increases both remanence and coercivity of magnetic materials. In contrast to previously known methods where magnetic moment of each particle was assumed to be 'fixed' with respect to the particle itself, our approach takes into account the (field-dependent) deviation of this moment from the particle anisotropy axis that occurs even for magnetically 'hard' particles possessing a strong mechanical contact. We show, that this deviation leads to the existence of the optimal value of the applied field for which the particle orientation (or alignment) time is minimal. The influence of the external pressure and internal mechanical friction on the details of the compactization/orientation process is also studied.
\end{abstract}

\date{\today}

\section{Introduction}



Discrete particle simulation is an enormously powerful predictive tool widely used in both fundamental science and modern engineering, as well as in industrial applications \cite{PoschelBook2005, Clearly2000}. Present computational capabilities allow to understand the relationship between microstructure of materials and their macroscopic properties \cite{Zhu2007}, as well as to provide recommendations for the materials fabrication processes. In particular, simulations of the mechanical rotation of magnetic particles in a powder during the compaction process and its influence on the resulting properties of a magnetic nanocomposite is of special importance for the material development in permanent magnets technology.

It is well known that in a system of non-interacting single-domain uniaxial magnetic particles their perfect alignment doubles the remanence and coercivity (compared to the system of particles with randomly oriented anisotropy axes). The corresponding state in nanocomposite magnetic materials is usually achieved by alignment of anisotropy axes of all crystallites using the mechanical rotation of particles. In real systems, the interparticle interaction and multidomain configurations decrease the effect, but it is still significant, so that the optimization of this alignment process is an important technological task.

For this purpose we developed a numerical algorithm for simulations of the mechanical rotation of magnetic particles in a powder under the influence of an external magnetic field and applied pressure. Our simulations are intended to predict the dependence of the particle orientation time on the field magnitude used during the orientation process and on the structural and magnetic material parameters, such as the particle size distribution, particle anisotropy constant and magnetization. All numerical studies of the process of magnetic-assisted compaction of hard magnetic particles conducted up to now (see, e.g., simulations of Nd–Fe–B powders \cite{Golovina2014,Soda2015}) have explicitly or implicitly assumed that the particle magnetic moment is aligned along the particle anisotropy axis, so that the torque from external field is effectively applied on the particle itself. However, this assumption is violated in sufficiently strong magnetic fields even for the rare-earth magnets, not to mention ferrite-based magnets, where deviations of magnetic moment from the anisotropy axis direction can be large already in the fields $\sim 1$ kOe.

Hence, the algorithm for realistic simulations of the compaction/alignment process should take into account (i) mechanical forces and torques arising by contact with other particles and surrounding box sides, (ii) external field and internal anisotropy torques acting on particle magnetic moment and (iii) mechanical torque acting on a particle when its magnetic moment is deflected from the anisotropy axis direction. Below, we describe the corresponding formalism and analyze physical results obtained using our algorithm.

\section{Model description}
\label{sec:model}

The proposed algorithm consists of the following stages:
(a) definition of system parameters, such as the number of particles, their size distribution and magnetic parameters (magnetization and anisotropy constant);
(b) generation of the close packing of particles (Sec. \ref{sec:gen_non-overlap});
(c) initialization of translational  $\{\vec{v}_i\}$ and rotational $\{\vec{\omega}_i\}$ particle velocities;
(d) calculation of the magnetic moments directions for all particles (Sec. \ref{sec:magnetic_part}) from the current direction of their anisotropy axes $\vec{n}_i$ and the value of applied magnetic field;
(e) numerical solution of equations of motion (Sec. \ref{sec:mech_part} ) with respect to $\vec{v}_i$ and $\vec{\omega}_i$, whereby spatial coordinates, anisotropy axes directions and directions of particles magnetic moments  are updated.

At present, the algorithm works with single-domain magnetic particles having a spherical shape; generalization of our method towards shape-anisotropic particles will be considered elsewhere.

\subsection{Generation of a collection of non-overlapping particles}
\label{sec:gen_non-overlap}

First, we create a system of non-overlapping spheres with prescribed size distribution within a rectangular box; this system should mimic a magnetic powder within a container having the corresponding form. We are able to generated a large number of non-overlapping spheres (up to $10^6$) using a method, which simulates a system of spheres interacting via a short-range repulsive potential described in details in our paper {\cite{Michels2014JMMM}}. In this algorithm, sphere centers are positioned randomly at the beginning of simulations. Then, spheres are moved according to the purely dissipative (i.e., neglecting the inertial term) equation of motion resulting from this potential. Due to its repulsive nature, the motion of spheres leads to a decrease of their mutual overlaps. We stop this procedure when the last overlap vanishes. After the spheres have been positioned, the initial velocities $\vec{v}_i$ and  $\vec{\omega}_i$ are initialized.

\subsection{Evaluation of the magnetic torque}
\label{sec:magnetic_part}

We use the Stoner-Wohlfarth model {\cite{Stoner1948}} to describe the behaviour of the particle magnetic moment in an external field. This model is supposed to correctly describe monocrystalline magnetic particles made of 'hard' magnetic materials (i.e., materials with the high magnetic anisotropy) which are normally used for producing permanent magnets. The restriction of the particle monocrystallinity can be omitted, if one uses the effective magnetic anisotropy (obtained after the averaging over anisotropy axis directions and anisotropy constants of crystallites constituting a polycrystalline magnetic particle) instead of the nominal magnetic anisotropy of the monocrystalline bulk material.

\begin{figure}[H] 
        \centering \includegraphics[width=0.6\columnwidth]{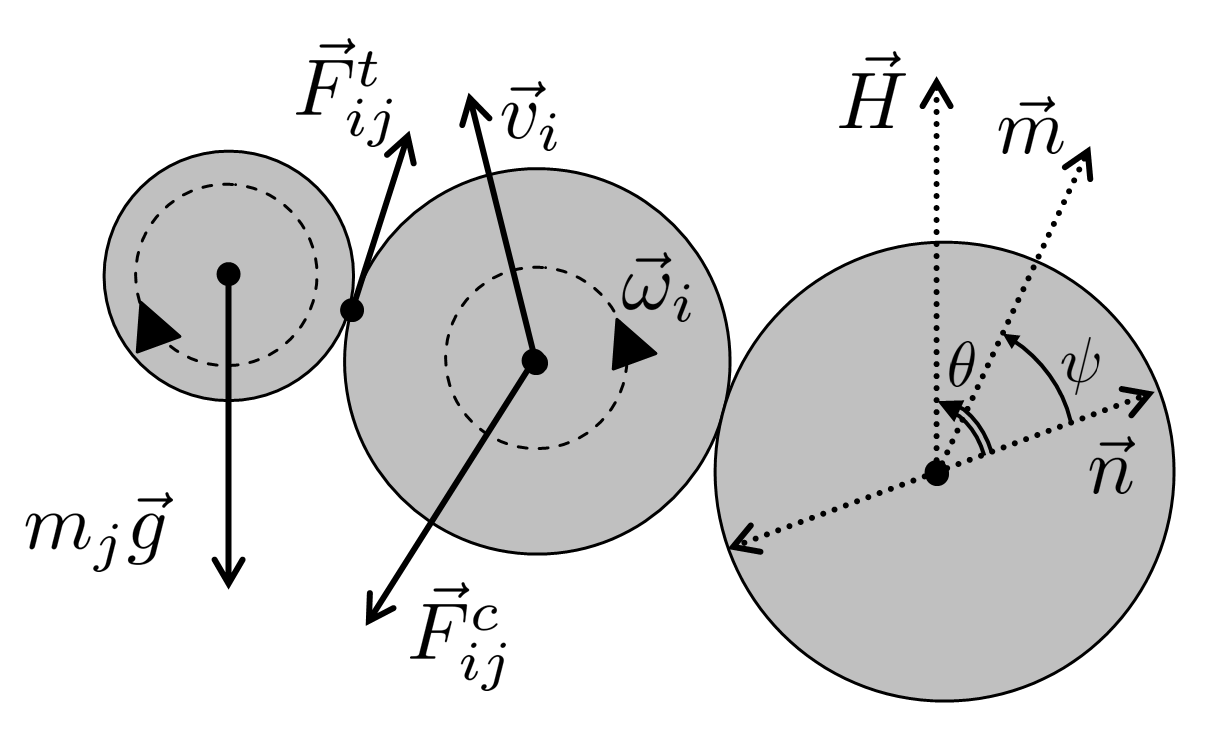}
        \caption{\label{fig:geometry2D} 2D representation of the system geometry, mechanical forces and magnetic quantities.}
\end{figure}

In the corresponding model, magnetic energy $E$ of the particle with the volume $V$, saturation magnetization $M_s$ and uniaxial anisotropy constant $K$ is the sum of the energy in the external field $H$ and the anisotropy energy

\begin{equation} \label{eq:magnetic_energy}
E = -M_s V H \cos (\theta-\psi) - K V \cos^2 \psi,
\end{equation}

where $\theta$ is the angle between the direction of the external field $\vec{H}$ and the anisotropy axis of the particle $\vec{n}$; $\psi$ is the angle between the particle magnetic moment $\vec{m}$ and the same anisotropy axis (see Fig. \ref{fig:geometry2D}). 

Angle $\psi$ as the function of the external field magnitude $H$ and the anisotropy axis orientation $\theta$ can be found from the equilibrium equation for the magnetic moment ($\delta E/\delta \psi = 0$) \cite{Stoner1948}:

\begin{equation} \label{eq:magnetic_equilibrium}
\frac{H}{\beta M_s} \sin(\theta-\psi) - \frac{1}{2} \sin 2\psi = 0,
\end{equation}

where the dimensionless anisotropy constant $\beta = 2K/M_s^2$ (which roughly describes the ratio between the anisotropy and magnetodipolar interparticle interaction energy in a magnetic particle system) has been introduced. 

Eq. (\ref{eq:magnetic_equilibrium}) should be solved numerically for all particle orientation angles $0 \leq \theta \leq \pi/2$ and each particular set of magnetic parameters including the magnitude of the external field.

\begin{figure}[H] 
        \centering \includegraphics[width=0.5\columnwidth]{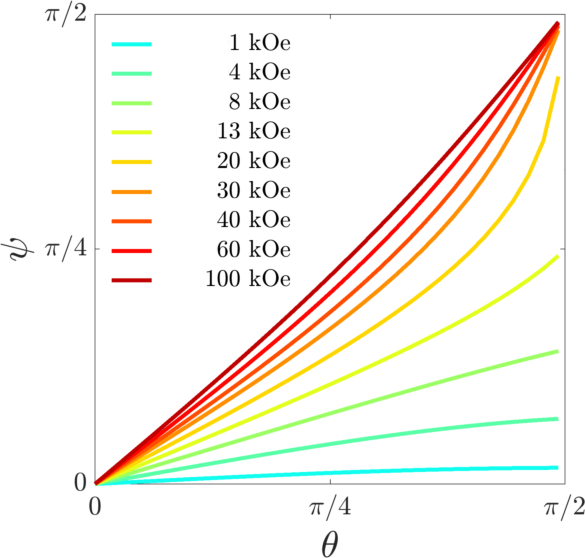}
        \caption{\label{fig:magnthetapsi} Dependence of the angle $\psi$ between the magnetic moment $\vec{m}$ and anisotropy direction $\vec{n}$ on the angle $\theta$ (between $\vec{n}$ and $\vec{H}$) for various values of the external magnetic field $H$.}
\end{figure}

A set of $\psi(\theta)$-dependencies for several values of the external field is presented in Fig. \ref{fig:magnthetapsi}, where magnetic parameters of the particle material $M_s = 380 \, {\rm G}$ and $K = 3.5 \cdot 10^6 \, {\rm erg/cm^3}$, corresponding to material properties of Sr-ferrite were used. The reduced anisotropy $\beta = 2K/M_s^2 \approx 50$ for this material is very large, indicating that the influence of the magnetodipolar interaction is extremely small. Due to this feature we have neglected this interaction in our simulations, thus greatly reducing the computational effort.

Having the $\psi(\theta, H)$-dependencies to our disposal, we can proceed with the evaluation of the torque which leads to the mechanical rotation of a magnetic particle, causing the alignment of all particle anisotropy axes along the applied field. It is important to understand, that in this model the torque $\vec{T}^{\rm mag}$ acting on a particle is {\it not} equal to the torque acting from the external field on the magnetic moment (fixed along the anisotropy axis, as assumed in other papers on this topic). Instead, the mechanical torque acting on a particle arises due to the {\it deviation} of its magnetic moment from the anisotropy axis direction due to the presence of the magnetocrystalline anisotropy. Hence this torque should evaluated as $\vec{T}^{\rm mag} = - \vec{\mu} \times \vec{H}_{\rm an}$, where the anisotropy field is given by $\vec{H}_{\rm an} = (2K/M_s)(\vec{m} \cdot \vec{n})\vec{n}$ (see, e.g., \cite{Stoner1948}). Thus, the resulting expression for the magnetic torque is 

\begin{equation} \label{eq:vector_torque_estimation}
\vec{T}^{\rm magn} = - \vec{\mu} \times \vec{H}_{\rm an} = 
-2KV(\vec m \cdot \vec n) [\vec m \times \vec n] 
\end{equation}

From this consideration it is clear that $T^{\rm mag}$ is proportional to the anisotropy field $H_{\rm an}$ and - due to the presence of the vector product $\vec{\mu} \times \vec{H}_{\rm an}$ - to the sine of the angle between $\psi$ between the anisotropy field $\vec{H}_{\rm an} \parallel \vec{n}$ and the magnetic moment direction $\vec{m}$. In turn, the anisotropy field value is $H_{\rm an}=H_k \cos{\psi}$, where $H_k = 2K/M_s$, so that the magnetic torque finally becomes

\begin{equation} \label{eq:torque_estimation}
T^{\rm mag} = \mu H_{\rm an} \sin{\psi} = \mu H_k \cos{\psi} \sin{\psi} = (\mu H_k/2) \sin{2\psi}
\end{equation}

\begin{figure}[H] 
        \centering \includegraphics[width=1.0\columnwidth]{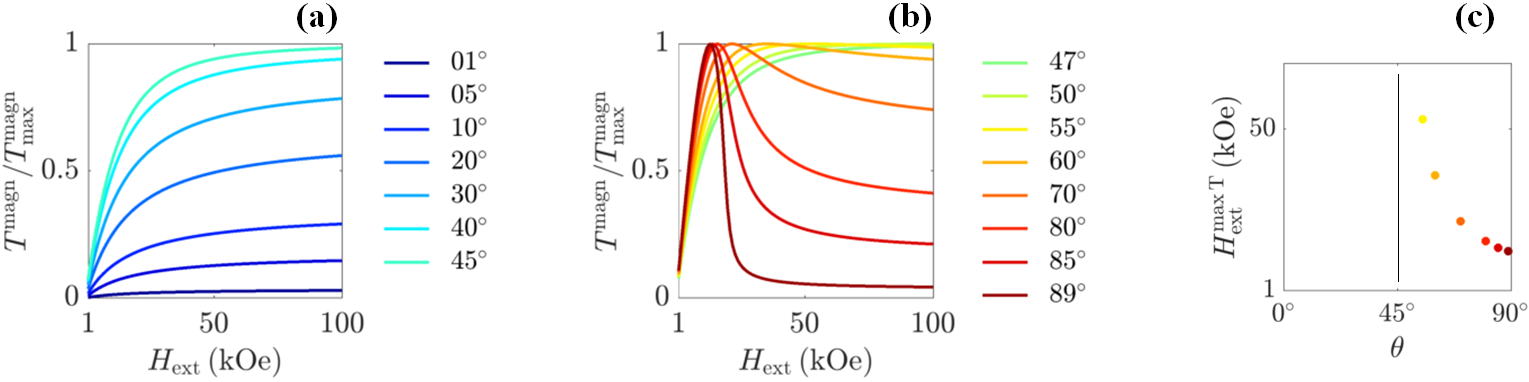}
        \caption{\label{fig:magnsin2psiHext} (a,b) Dependencies of the torque magnitude $T^{\rm mag} \propto  \sin{2 \psi}$ on the external magnetic field $H_{\rm ext}$ for various orientations of the anisotropy axes $\theta$. (c) External field values corresponding to the \textit{maximal} torque magnitude $T^{\rm mag}$ as the function of the angle $\theta$ for $\theta > 45^{\circ}$.}
\end{figure}

Solving Eq. (\ref{eq:magnetic_equilibrium}) numerically for each value of the angle $\theta$ and the external field $H_{\rm ext}$, we obtain the dependence $\psi(H_{\rm ext})$, which is then substituted into Eq. (\ref{eq:torque_estimation}). The result is shown in Fig. \ref{fig:magnsin2psiHext}, where the dependence of the magnetic torque on the external field for various values orientations of the anisotropy axis $\theta$ is displayed. The most important feature of these results is the presence of a maximum on   the $T^{\rm magn}(H_{\rm ext})$-dependence for $\theta > 45^{\circ}$, which becomes sharper with increasing $\theta$. External magnetic field which delivers the maximal torque magnitude is plotted as a function of $\theta$ in Fig. \ref{fig:magnsin2psiHext}(c). For $\theta \geqslant 45^{\circ}$ corresponding field value represents the optimal magnitude of the external magnetic field for the alignment of particles which anisotropy axes have the given orientation $\theta$ with respect to the applied field $\vec H$.

\subsection{Mechanichal part}
\label{sec:mech_part}

In this section we present the mechanical model used in our calculations, providing  equations of the translational and rotational particles motion. We have chosen a relatively simple mechanical model; a more sophisticated one can be used if necessary, employing the same magnetic part described in the previous section. 

We consider the dynamics of mechanically interacted spherical particles governed by the following set of equations \cite{Luding1998,PoschelBook2005,Schwager2008}: 

\begin{equation} 
        m_i \dot{\vec{v}}_i = \sum\limits_{j}\vec{F}_{ij}^c + m_i\vec{g} - \gamma \vec{v_i}, \label{eq:motion_force}
\end{equation}

\begin{equation} 
        I_i \dot{\vec{\omega}}_i =  \sum\limits_{j} \vec{R}_i \times \vec{F}^t_{ij} + \vec{T}_i^{\rm magn}   - R_i^2\gamma^\star \vec{\omega_i}\label{eq:motion_torque}
\end{equation}
Here $m_i$, $R_i$, $\vec{v}_i$, $\vec{\omega}_i$ denote the mass, radius, translational and angular velocity of the $i$-th particle correspondingly, and $\vec{g}$ is the gravitational acceleration. The last terms of both equations  introduce the viscoelastic damping into the system with the normal $\gamma$ and tangential $\gamma^\star$ contact constants.

In the description of the translational motion, the contact force $\vec{F_{ij}^c}$ between $i$-th and $j$-th particles in the Eq. (\ref{eq:motion_force}) is a sum of normal $\vec{F}^n_{ij}$ and tangential $\vec{F}^t_{ij}$ components defined as

\begin{equation}
       \vec{F}^n_{ij} = -\frac{4}{3}E_{ij}^\star \vec{\delta}_n 
       \sqrt{\delta_n R_{ij}^\star}, 
    \label{eq:force_normal}
\end{equation}
\begin{equation}
    \vec{F}^t_{ij} = - \frac{\vec{v}^t_{\rm rel}}{v^t_{\rm rel}} \mu F^n_{ij}  \label{eq:force_tangential}
\end{equation} 

where $\vec{\delta}_n = \vec{R}_i + \vec{R}_j - \vec{d}_{ij}$ is the relative translational displacement of spherical particles $i$ and $j$ and $\vec{v}^{\,t}_{\rm rel} = \vec{v}_i-\vec{v}_j-(R_i \vec{\omega}_i+R_j \vec{\omega}_j) \times \vec{e}^{\,n}_{ij}$ is a total relative velocity resulting from the motion of the sphere's centers and from the sphere's rotation. Reduced radii and Young's moduli are computed as

\begin{equation} \label{eq:reducedRE}
\frac{1}{R_{ij}^\star} = \frac{1}{R_i} + \frac{1}{R_j}, \,\, {\rm and} \,\, \frac{1}{E_{ij}^\star} = \frac{1-\nu^2_i}{E_i} + \frac{1-\nu^2_j}{E_j}
\end{equation}

correspondingly ($\nu$ is the Poisson ratio). Eqs. (\ref{eq:force_normal}) means that for the description of the mechanical contact we have chosen the elastic spheres contact approximation (Hertz theory) \cite{Johnson1985}. Tangential friction employed in Eq. (\ref{eq:force_tangential}) depends on the friction coefficient $\mu$, direction of the relative velocity $\vec{v}^t_{\rm rel}$ and the magnitude of normal contact force $F^n_{ij}$ (see Eq. (\ref{eq:force_normal})).

Additional contact forces due to the impact between particles and box sides are described analogously to the particle-particle interaction:  $\vec{v}_j$ and $\vec{\omega}_j$ in the formula for $\vec{v}^{\,t}_{\rm rel}$ are substituted by zero and $\vec{e}^{\,n}_{ij}$ is a vector normal to the corresponding box side plane.

For the rotational motion, the torque terms in Eq. (\ref{eq:motion_torque}) result from mechanical tangential forces due to the contact with other particles and from the magnetic field; the evaluation receipt for the latter term $\vec{T}_i^{\rm mag}$ is provided in the Sec. \ref{sec:magnetic_part}. The moment of inertia of the $i$-th particle is $I_i = (2/5)m_iR_i^2$ for particles having the spherical shape. In our simulations we use elastic constants $E = 1.8 \cdot 10^{11} \, \mathrm{Pa}$ and $\nu = 0.28$ which are typical for ferrites. 

The total number of particles (see Sec. \ref{sec:results}) is $N = 1000$, which diameters are generated according to the log-normal distribution with $\sigma = 0.10$ and the average diameter $D_{\mathrm{av}}  = 1 \, \mathrm{\mu m}$.

We solve Eqs. (\ref{eq:motion_force}) and (\ref{eq:motion_torque}) neglecting the inertial terms. This approximation can be easily justified for the compactization and alignment process under consideration using standard estimations similar to those employed in the particle dynamics in a viscous fluid. In our case, an additional advantage of this approach is the absence of small bouncing movements of particles (present when solving full equations of motion) which would not change the final result, but would lead to a very slow convergence of the numerical procedure.

Heun's method \cite{HeunMethod} was applied to solve the corresponding set of ordinary differential equations corresponding to translational and rotational motions. For the description of 3D rotations we use a quaternion technique, which is briefly outlined in Appendix \ref{appendix:rot3D}. Reduced units employed for the stability of the numerical scheme are described in Appendix \ref{appendix:reduced_units}. 

\section{Results and discussion}
\label{sec:results}
\begin{figure}[H] 
        \centering \includegraphics[width=0.95\columnwidth]{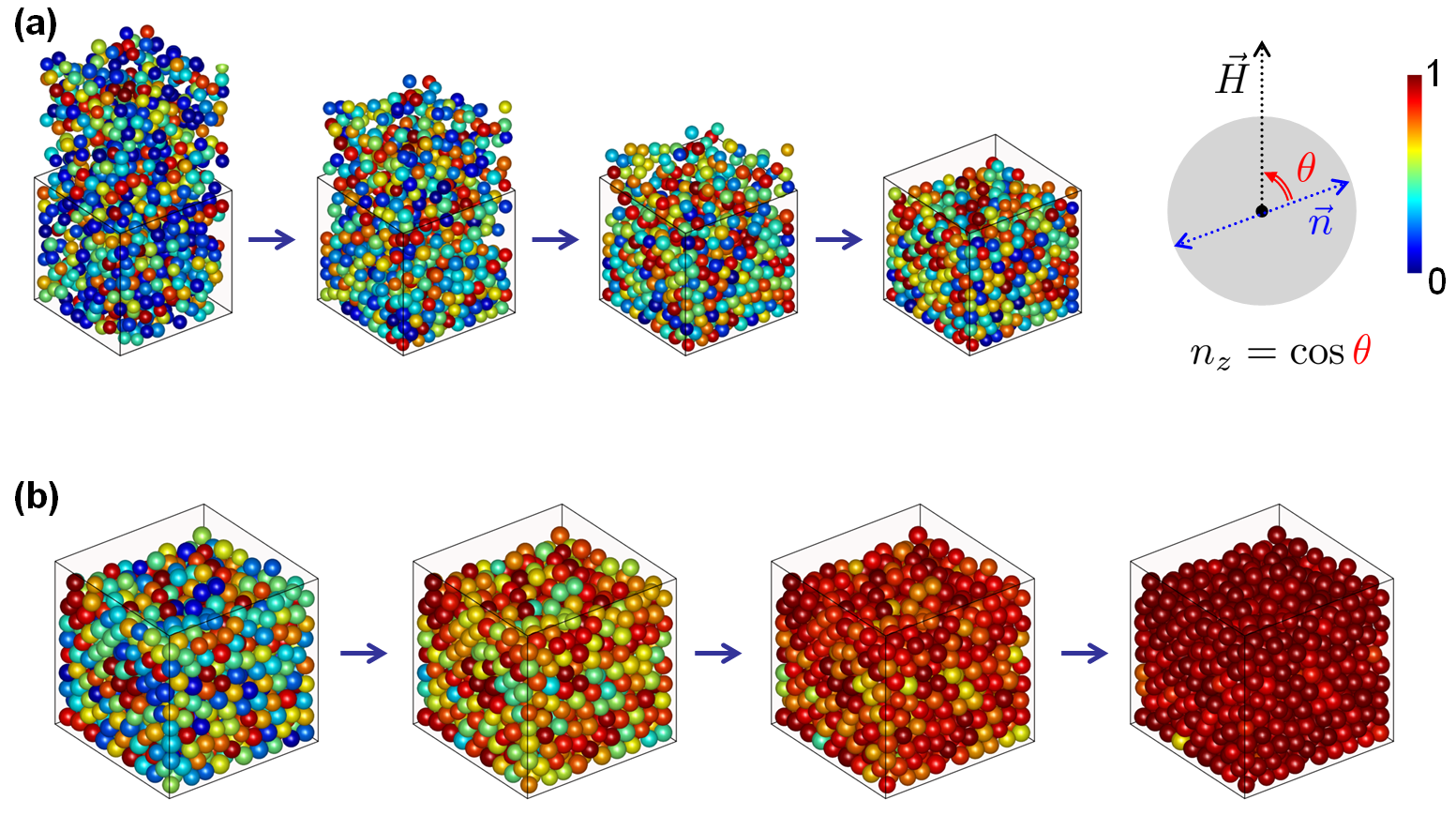}
        \caption{\label{fig:compact3Dcolor} (a) Consecutive system states during the initial stage of the compactization process (particles are 'dropped' into the compactization box without external magnetic field). (b) System compactization and alignment \textbf{in presence} of an external magnetic field: consecutive  system states during the compaction under pressure. Color of particles corresponds to the projection of particle anisotropy axes on the magnetic field direction (vertical axis): dark red - along the field, blue - in the perpendicular direction.}
\end{figure}

In this section we present physical results obtained by applying our algorithm to a system of magnetic particles with parameters listed in Sec. \ref{sec:model}. Every simulation described below consists of three stages: (i) particles are positioned randomly within an initial box which $x$- and $y$-sizes are equal to the corresponding sizes of the container where the alignment and compactization will occur; the vertical size of this initial box is 3 times larger than the size of the compactization container. (ii) particle overlaps are eliminated as described in Sec. 2.1 and then  particles are 'dropped' into the compactization container without external pressure and applied magnetic field (Fig. \ref{fig:compact3Dcolor}a), but taking into account the gravitational acceleration; (iii) particles are aligned under the influences of the external magnetic field and an optional pressure (Fig. \ref{fig:compact3Dcolor}b) by numerically solving the equations of motion described in the previous section.

\begin{figure}[H] 
        \centering \includegraphics[width=0.7\columnwidth]{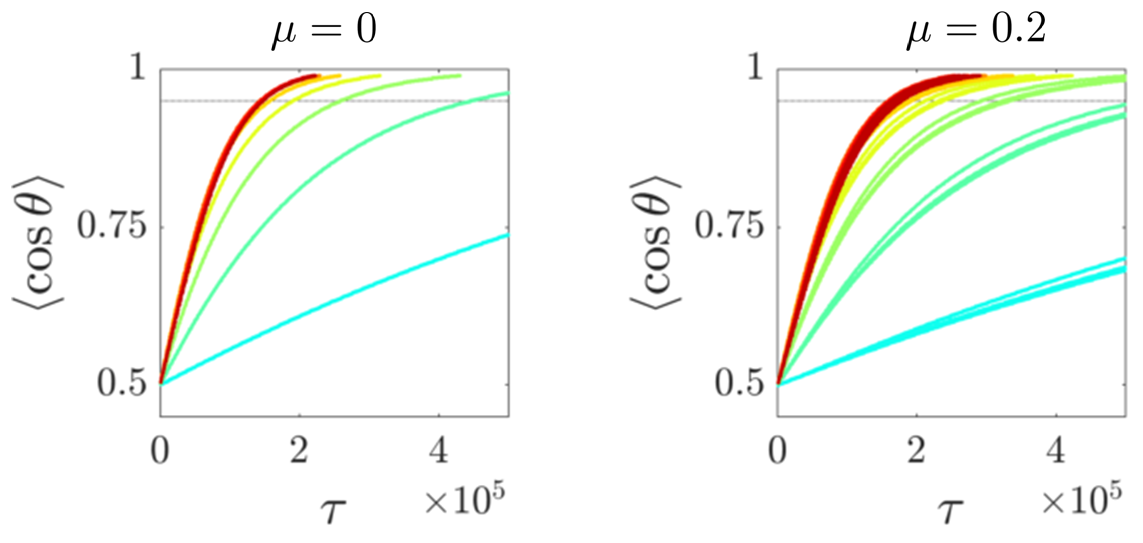}
        \caption{\label{fig:compact_mu0_mu02} Numerically obtained time dependencies of the alignment degree of particle systems with various friction coefficients $\mu$ under the influence of external field. Color legend corresponding to different magnetic field magnitudes is the same as at Fig. \ref{fig:magnthetapsi}. Lines of the same color demonstrate the alignment of different geometrical realizations of the system with the same parameter set. Thin horizontal grey line corresponds to $\langle\cos{\theta}\rangle = 0.95$.}
\end{figure}

Fig. \ref{fig:compact_mu0_mu02} demonstrates two sets of results of this procedure at the pressure of $0.5 \, \mathrm{atm}$ with different magnitudes of magnetic field applied for the system without internal friction (left panel) and with friction coefficient $\mu = 0.2$ (right panel). The first obvious conclusion of this comparison is that the alignment occurs faster in the system with smaller friction (note that the alignment time is not zero for $\mu = 0$ due to the presence of viscoelastic terms in equations of motion (\ref{eq:motion_force}) and (\ref{eq:motion_torque})). Second, we point out that for both cases the alignment time almost does not change starting from $20 \, \mathrm{kOe}$ and that this time at $100 \, \mathrm{kOe}$ is even larger than the corresponding value at $40 \, \mathrm{kOe}$.

\begin{figure}[H] 
        \centering \includegraphics[width=0.95\columnwidth]{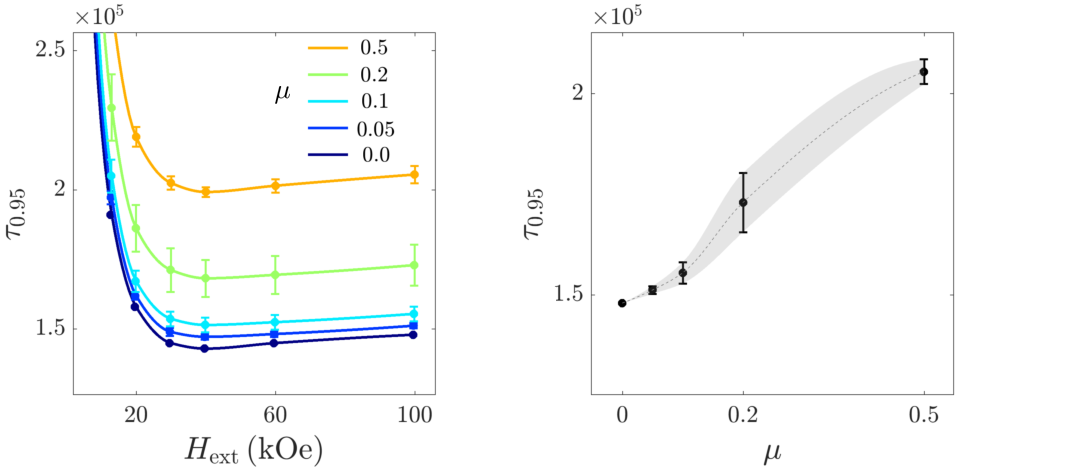}
        \caption{\label{fig:set_mu} (a) Alignment time required to achieve the orientation degree $\langle\cos{\theta}\rangle = 0.95$ (see text for details) as the function of the applied field for various internal friction coefficients $\mu$. (b) Dependence of the alignment time on the friction coefficient at $H = 40 \, \mathrm{kOe}$. Dashed line is the guide for an eye. Shaded area represent standard deviation of the quantity of interest obtained for different geometrical realizations.}
\end{figure}

The detailed analysis of this behaviour is presented on Fig. \ref{fig:set_mu}, where the time of $95\%$ alignment (the time moment when $\langle\cos{\theta}\rangle = 0.95$) is plotted vs the external field magnitude. These simulations have been performed for different internal friction coefficients $\mu$ demonstrating that the minimum of this dependence is achieved at $H \approx 40 \, \mathrm{kOe}$ for all friction values. This observation specifies the optimal external field for the alignment of this type of material. Exceeding this optimal field does not provide any advantage, leading even to a little increase of the orientation time. We emphasize that the physical reason for the existence of the optimal alignment time is the non-monotonic dependence of the magnetic torque on the external field value for particles with $\theta > 45^{\circ}$ discussed in the section \ref{sec:magnetic_part}.

The nonlinearity of the alignment time as the function of the friction coefficient (shown in Fig. \ref{fig:set_mu}b for $H = 40 \, \mathrm{kOe}$ is due to the nonlinear dependence of the mechanical contact force on the translational displacement used in the model, which affects the tangential part of this force. The increase of the friction coefficient from $0.0$ to $0.5$ results in the alignment time increase of $35\%$.

\begin{figure}[H] 
        \centering \includegraphics[width=0.9\columnwidth]{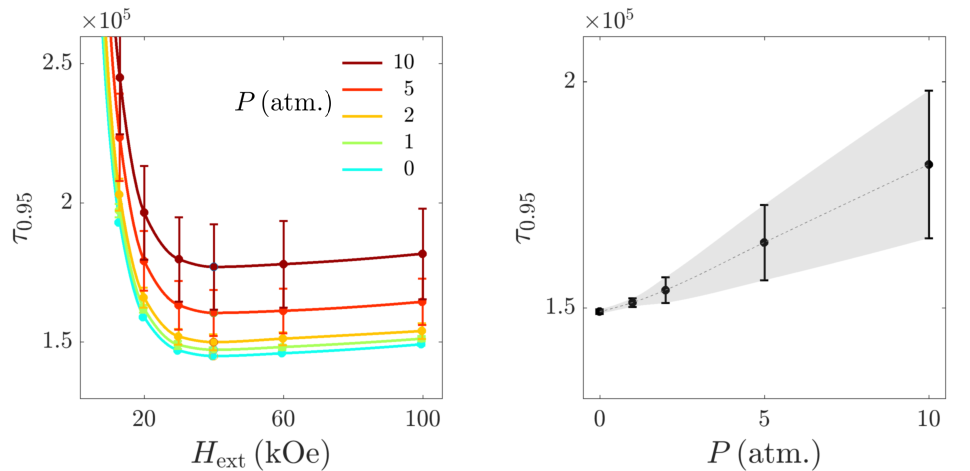}
        \caption{\label{fig:set_pressure} (a) Alignment time $\tau_{0.95}$ as the function of the external field for various values of the applied pressure $P$. (b) Dependence of the alignment time on the applied pressure at the external field $H = 40 \, \mathrm{kOe}$ shown in the same manner as in the previous figure.}
\end{figure}

 Next, we have studied the effect of the external pressure which is often used by the powder compactization. To simulate this pressure, we have applied an extra force (aligned with the gravitational acceleration) to all particles in the upper particle layer, i.e. to particles which $z$-coordinate was $z_i > 0.9 z_{\rm max}$. To model the constant pressure, this force was taken to be proportional to $R_i^2$. The results of these simulations showing the dependence of the alignment time on the external pressure (with $\mu = 0.05$) are presented in Fig. \ref{fig:set_pressure}. Raising the pressure up to $10 \, \mathrm{atm}$ increases the time of the alignment process only by approximately $25\%$.

One of the most interesting features of results shown in Figs. \ref{fig:set_mu} and  \ref{fig:set_pressure} is the fact that the position of the minimum is nearly independent on the friction coefficient and external pressure, i.e., this position is the universal characteristic of the system under study. This statement might be especially useful in the case when experimental data on the friction coefficient are unavailable.

\section{Conclusion}
 In this paper, we have presented a numerical method for modeling the compaction and mechanical orientation process of magnetic powders. In contrast to previously published approaches, our method takes into account the deviation of the particle magnetic moment from the particle anisotropy axis. The presented method has been applied to a system of hard ferrite particles. Our studies of the influence of the external magnetic field on the duration of the alignment process have demonstrated that there exist an optimal value of this field (around 40 kOe for Sr-ferrite), above which the alignment time increases. This qualitatively new feature is due to the non-monotonic dependence of the magnetic torque acting on the particle on the external field value. Further, the influence of the external mechanical pressure and surface coating of magnetic particle (e.g. materials which reduces the interparticle friction and can be removed after the orientation process) has been studied. The presented methodology can be used to provide recommendations for the optimal compaction and alignment conditions of magnetic powders used in production of permanent magnets.

\section{Acknowledgement}
Authors gratefully acknowledge financial support of EU-H2020 AMPHIBIAN Project (720853).

\bibliographystyle{apalike}
\bibliography{particles_alignment}

\appendix
\section{Rotations in 3D}
\label{appendix:rot3D}
All simulation techniques where 3D rotations are involved require a method to describe the spatial and angular positions of objects under study (magnetic particles in our case). One of the most effective approaches is the usage of quaternion algebra~ \cite{wiki:quaternion} which extends the formalism of complex numbers. A quaternion can be presented in the form
\begin{equation} \label{eq:quaternion_def1}
\vec{q} = s+p_1 \ \mathbf {i} +p_2 \ \mathbf {j} +p_3 \ \mathbf {k}, 
\end{equation}
where $s$, $p_1$, $p_2$, and $p_3$ are real numbers, and $\mathbf {i}$, $ \mathbf {j}$, and $\mathbf {k}$ are the fundamental quaternion units. Depending on the application, quaternions can be used together with other methods, such as Euler angles and rotation matrices, or as an alternative to them. One of the main advantages of quaternion algebra is that the rotation axis $\vec{\phi}$ and angle of rotation $\phi$ can be easily recovered:
\begin{equation}
    \vec{q} = [s,\vec{p}] = 
    \left[\cos \frac{\phi}{2}, \sin\frac{\phi}{2}\vec{\phi}\right],
    \label{eq:quaternion_def2} 
\end{equation} 
    
\begin{equation}
      ||q||^2 = s^2 + ||p||^2 = 1.  \label{eq:quaternion_def3}
\end{equation}
The noncommutative multiplication of quaternions corresponds to two consecutive rotations is a more concise representation than a rotation matrix:
\begin{equation}
    \label{eq:quaternion_mult}
        \tilde{s} = s \cdot \delta s - \vec{p} \cdot \vec{\delta p},
\end{equation}
\begin{equation}
        \tilde{p} = s \cdot \delta p + \delta s \cdot p - \vec{p} \times \vec{\delta p}.
\end{equation}
In such a way 3D rotations of particles can be easily composed having less multiplication and addition operations and in a more concise representation than for a rotation matrix \cite{arxiv:Graf2008}.

\section{Reduced units used in simulations}
\label{appendix:reduced_units}

To obtain dimensionless geometrical parameters, we have chosen the average particle radius $R_{\mathrm{av}}$. so that $\tilde{\delta} = \delta / R_{\mathrm{av}}$, $\tilde{R} = R / R_{\mathrm{av}}$ and $\vec{\tilde{r}} = \vec{r}/ R_{\mathrm{av}}$. Thus, the Eqs. \ref{eq:motion_force} and \ref{eq:motion_torque} after neglecting the inertial terms (as explained in the main text) can be rewritten as
 
 \begin{equation} \label{eq:noninertial_force_reduced}
 \frac{ d\vec{\tilde{r}}}{d\tau} = 
-\sum\limits_{c}\tilde{E} \vec{\tilde{\delta}} \sqrt{\tilde{\delta} \tilde{R}^\star} \vec{e}_n
-\sum\limits_{c} \mu \tilde{E} \vec{\tilde{\delta}} \sqrt{\tilde{\delta} \tilde{R}^\star} \vec{e}_t
-\tilde{g}\tilde{R}^3\vec{e}_z - \tilde{p}_z \tilde{R}^2\vec{e}_z 
\end{equation}

\begin{equation} \label{eq:noninertial_moment_reduced}
 \frac{\gamma^\star}{\gamma} \tilde{R}^2 \frac{ d\vec{\tilde{\phi}}}{d\tau} = 
\sum\limits_{c}
\vec{\tilde{R}} \times \mu (-\tilde{E}^\star \vec{\tilde{\delta}} \sqrt{\tilde{\delta} \tilde{R}^\star} ) \vec{e}_t
- \tilde{K} \tilde{R}^3 (\vec m \cdot \vec n) [\vec m \times \vec n] 
\end{equation}
 
with reduced units $\tau = t (E R_{\mathrm {av}}) / \gamma$, $\tilde{g} = (4 \pi/3) 
(\rho g R_{\mathrm {av}}/E)$, $\tilde{p}_z = \pi P_z/E$ and $\tilde{K} = 8\pi K/ 3 E$. For simplicity, normal and tangential viscoelastic damping constants are chosen to be equal ($ \gamma^\star = \gamma$) with $\gamma = 20 \, \mathrm{ N \cdot s/m}$, leading to the characterised time $t_{\mathrm{char}} = \gamma/(E R_{\mathrm {av}}) = 2.2 \cdot 10^{-4} \, \mathrm s$.

\end{document}